\documentclass[journal]{IEEEtran}

\usepackage{cite,graphicx,subfigure,hyperref,amsmath,amsthm,amsfonts,array,fp} 
 
\usepackage{pict2e}

\usepackage{pgfplots}

\pgfplotsset{compat=newest}
\pgfplotsset{plot coordinates/math parser=false}
\newlength\figureheight
\newlength\figurewidth

\newlength{\Lpr}
\newsavebox{\Bpr}

\newcommand{\D}[1]{\ensuremath{\displaystyle #1}}

\newcommand{\V}[1]{\mbox{\boldmath$#1$\unboldmath}}

\newcommand{\bdm}{\begin{displaymath}}
\newcommand{\edm}{\end{displaymath}}

\newcommand{\be}[1]{\begin{equation} \label{#1}}
\newcommand{\ee}{\end{equation}}

\newcommand{\bae}[3]{
\begin{equation} \label{#1}
\renewcommand{\arraystretch}{#2}
\begin{array}{#3}}

\newcommand{\eae}{\end{array}\end{equation}}

\newcommand{\baen}[2]{
\begin{displaymath} 
\renewcommand{\arraystretch}{#1}
\begin{array}{#2}}

\newcommand{\eaen}{\end{array}\end{displaymath}}

\newcommand{\DefLetter}[4]{
\newcommand{#1}{\ensuremath{\V{#2}}} 
\newcommand{#3}{\ensuremath{\V{#4}}} 
}

\DefLetter{\vzer}{0}{\mzer}{0}
\DefLetter{\vone}{1}{\mone}{1}
\DefLetter{\va}{a}{\ma}{A}
\DefLetter{\vb}{b}{\mb}{B}
\DefLetter{\vc}{c}{\mc}{C}
\DefLetter{\vd}{d}{\md}{D}
\DefLetter{\ve}{e}{\me}{E}
\DefLetter{\vf}{f}{\mf}{F}
\DefLetter{\vg}{g}{\mg}{G}
\DefLetter{\vh}{h}{\mh}{H}
\DefLetter{\vi}{i}{\mi}{I}
\DefLetter{\vj}{j}{\mj}{J}
\DefLetter{\vk}{k}{\mk}{K}
\DefLetter{\vl}{l}{\ml}{L}
\DefLetter{\vm}{m}{\mm}{M}
\DefLetter{\vn}{n}{\mn}{N}
\DefLetter{\vpr}{p}{\mpr}{P}
\DefLetter{\vq}{q}{\mq}{Q}
\DefLetter{\vr}{r}{\mr}{R}
\DefLetter{\vs}{s}{\ms}{S}
\DefLetter{\vt}{t}{\mt}{T}
\DefLetter{\vur}{u}{\mur}{U}
\DefLetter{\vv}{v}{\mv}{V}
\DefLetter{\vw}{w}{\mw}{W}
\DefLetter{\vx}{x}{\mx}{X}
\DefLetter{\vy}{y}{\my}{Y}
\DefLetter{\vz}{z}{\mz}{Z}

\DefLetter{\vdel}{\delta}{\mdel}{\Delta}
\DefLetter{\vphi}{\phi}{\mphi}{\Phi}
\DefLetter{\vpsi}{\psi}{\mpsi}{\Psi}
\DefLetter{\vrho}{\rho}{\mrho}{\Lambda}
\DefLetter{\vxi}{\xi}{\mxi}{\Xi}

\DefLetter{\valpha}{\alpha}{\malpha}{\Alpha}
\DefLetter{\vbeta}{\beta}{\mbeta}{\Beta}
\DefLetter{\vlam}{\lambda}{\mlam}{\Lambda}
\DefLetter{\vsig}{\sigma}{\msig}{\Sigma}
\DefLetter{\vtau}{\tau}{\mtau}{\tau}
\DefLetter{\vtheta}{\theta}{\mtheta}{\Theta}
\DefLetter{\vome}{\omega}{\mome}{\Omega}
\DefLetter{\vzero}{0}{\mzero}{0}
\DefLetter{\vgam}{\gamma}{\mgam}{\Gamma}
\DefLetter{\veps}{\epsilon}{\meps}{\Epsilon}
\DefLetter{\veta}{\eta}{\meta}{\Eta}

\newcommand{\DefFuncLetter}[2]{
\newcommand{#1}{\ensuremath{{#2}}} 
}

\DefFuncLetter{\Fzer}{0}
\DefFuncLetter{\Fa}{a}
\DefFuncLetter{\FA}{A}
\DefFuncLetter{\Fb}{b}
\DefFuncLetter{\Fc}{c}
\DefFuncLetter{\FC}{C}
\DefFuncLetter{\Fd}{d}
\DefFuncLetter{\Fe}{e}
\DefFuncLetter{\Ff}{f}
\DefFuncLetter{\Fg}{g}
\DefFuncLetter{\FG}{G}
\DefFuncLetter{\Fh}{h}
\DefFuncLetter{\FH}{H}
\DefFuncLetter{\Fi}{i}
\DefFuncLetter{\Fk}{k}
\DefFuncLetter{\Fl}{l}
\DefFuncLetter{\FL}{L}
\DefFuncLetter{\Fm}{m}
\DefFuncLetter{\Fn}{n}
\DefFuncLetter{\Fnr}{n}
\DefFuncLetter{\FN}{N}
\DefFuncLetter{\Fo}{o}
\DefFuncLetter{\FO}{O}
\DefFuncLetter{\Fpr}{p}
\DefFuncLetter{\FPr}{P}
\DefFuncLetter{\Fq}{q}
\DefFuncLetter{\Fr}{r}
\DefFuncLetter{\Fs}{s}
\DefFuncLetter{\FS}{S}
\DefFuncLetter{\Ft}{t}
\DefFuncLetter{\FT}{T}
\DefFuncLetter{\Fu}{u}
\DefFuncLetter{\FU}{U}
\DefFuncLetter{\Fv}{v}
\DefFuncLetter{\Fw}{w}
\DefFuncLetter{\FW}{W}
\DefFuncLetter{\Fx}{x}
\DefFuncLetter{\Fy}{y}
\DefFuncLetter{\FY}{Y}
\DefFuncLetter{\Fz}{z}
\DefFuncLetter{\FZ}{Z}
\DefFuncLetter{\Falp}{\alpha}
\DefFuncLetter{\Fbet}{\beta}
\DefFuncLetter{\FBet}{B}
\DefFuncLetter{\Fchi}{\chi}
\DefFuncLetter{\Fdel}{\delta}
\DefFuncLetter{\Fzet}{\zeta}
\DefFuncLetter{\FEps}{\Epsilon}
\DefFuncLetter{\Feta}{\eta}
\DefFuncLetter{\Fphi}{\phi}
\DefFuncLetter{\FPhi}{\Phi}
\DefFuncLetter{\Fpsi}{\psi}
\DefFuncLetter{\FPsi}{\Psi}
\DefFuncLetter{\Fgam}{\gamma}
\DefFuncLetter{\FGam}{\Gamma}
\DefFuncLetter{\Flam}{\lambda}
\DefFuncLetter{\FLam}{\Lambda}
\DefFuncLetter{\Fsig}{\sigma}
\DefFuncLetter{\Ftau}{\tau}
\DefFuncLetter{\Fome}{\omega}
\DefFuncLetter{\Feps}{\epsilon}
\DefFuncLetter{\Fthe}{\theta}
\DefFuncLetter{\Fvar}{\vartheta}

\DefFuncLetter{\FB}{B}
\DefFuncLetter{\FD}{D}
\DefFuncLetter{\FE}{E}
\DefFuncLetter{\FF}{F}
\DefFuncLetter{\FI}{I}
\DefFuncLetter{\FJ}{J}
\DefFuncLetter{\FM}{M}
\DefFuncLetter{\FR}{R}
\DefFuncLetter{\FV}{V}
\DefFuncLetter{\FX}{X}

\newcommand{\DefCalLetter}[2]{
\newcommand{#1}{\ensuremath{\mathcal{#2}}} 
}
\DefCalLetter{\CC}{C}
\DefCalLetter{\CD}{D}

\DefCalLetter{\CS}{S}
\DefCalLetter{\CV}{V}

\newcommand{\DefSubLetter}[2]{
\newcommand{#1}{\mathrm{#2}} 
}

\DefSubLetter{\slzer}{0}
\DefSubLetter{\sla}{a}
\DefSubLetter{\slA}{A}
\DefSubLetter{\slb}{b}
\DefSubLetter{\slB}{B}
\DefSubLetter{\slc}{c}
\DefSubLetter{\slC}{C}
\DefSubLetter{\sld}{d}
\DefSubLetter{\slD}{D}
\DefSubLetter{\sle}{e}
\DefSubLetter{\slE}{E}
\DefSubLetter{\slf}{f}
\DefSubLetter{\slF}{F}
\DefSubLetter{\slg}{g}
\DefSubLetter{\slG}{G}
\DefSubLetter{\slh}{h}
\DefSubLetter{\slH}{H}
\DefSubLetter{\sli}{i}
\DefSubLetter{\slI}{I}
\DefSubLetter{\slk}{k}
\DefSubLetter{\sll}{l}
\DefSubLetter{\slL}{L}
\DefSubLetter{\slm}{m}
\DefSubLetter{\slM}{M}
\DefSubLetter{\sln}{n}
\DefSubLetter{\slnr}{n}
\DefSubLetter{\slN}{N}
\DefSubLetter{\slo}{o}
\DefSubLetter{\slp}{p}
\DefSubLetter{\slP}{P}
\DefSubLetter{\slq}{q}
\DefSubLetter{\slQ}{Q}
\DefSubLetter{\slr}{r}
\DefSubLetter{\slR}{R}
\DefSubLetter{\sls}{s}
\DefSubLetter{\slS}{S}
\DefSubLetter{\slt}{t}
\DefSubLetter{\slT}{T}
\DefSubLetter{\slu}{u}
\DefSubLetter{\slU}{U}
\DefSubLetter{\slv}{v}
\DefSubLetter{\slw}{w}
\DefSubLetter{\slW}{W}
\DefSubLetter{\slx}{x}
\DefSubLetter{\slX}{X}
\DefSubLetter{\sly}{y}
\DefSubLetter{\slY}{Y}
\DefSubLetter{\slz}{z}
\DefSubLetter{\slZ}{Z}

\DefSubLetter{\slalp}{\alpha}
\DefSubLetter{\slbet}{\beta}
\DefSubLetter{\sldel}{\delta}
\DefSubLetter{\slDel}{\Delta}
\DefSubLetter{\sleps}{\epsilon}
\DefSubLetter{\slgam}{\gamma}
\DefSubLetter{\slphi}{\phi}
\DefSubLetter{\sltau}{\tau}
\DefSubLetter{\slxi}{\xi}
\DefSubLetter{\slthe}{\theta}

\newcommand{\DFT}{\mathrm{DFT}}
\newcommand{\IDFT}{\mathrm{IDFT}}
\newcommand{\sump}{\sum_{p=1}^P}

\newcommand{\pE}[1]{\Fe^{j2\pi #1}}
\newcommand{\mE}[1]{\Fe^{-j2\pi #1}}
\newcommand{\SEQ}[3]{\{#1\}_{#2}^{#3}}
\newcommand{\Sp}{\SEQ{S_p} p P}
\newcommand{\sqa}{\SEQ{\Fs(q/P+ja)} q P}
\newcommand{\LK}{\SEQ{\FL(\Fe^{j2\pi (q/P+ja)})} q P}
\newcommand{\LS}{\Big\{\sum_{p=1}^P\frac{\Fs(t_p)}{\FL'(\Fe^{j2\pi t_p})(\Fe^{j2\pi
      (q/P+ja)}-\Fe^{j2\pi t_p})}\Big\}_q^P}
\newcommand{\LSb}{\Big\{\D\sum_{p=1}^P\frac{\Fs(t_p)\Fh(-Pt_p+Pja)}{\FL'(\Fe^{j2\pi
    t_p})\pE{t_p}}\Fh_1(q/P-t_p)\Big\}_q^P}
\newcommand{\Ld}{\SEQ{\FL'(\Fe^{j2\pi t_p})} p P}
\newcommand{\vseq}{\SEQ{\Fv(q/P)} q P}

\newcommand{\gridD}{
\linethickness{0.2mm}
\multiput(0,0)(0,1){11}{\line(1,0){10}}
\multiput(0,0)(1,0){11}{\line(0,1){10}}

\linethickness{0.05mm}
\multiput(0,0)(0,0.5){21}{\line(1,0){10}}
\multiput(0,0)(0.5,0){21}{\line(0,1){10}}

\linethickness{0.02mm}
\multiput(0,0)(0,0.1){101}{\line(1,0){10}}
\multiput(0,0)(0.1,0){101}{\line(0,1){10}}

\put(-0.09,-0.3){0} \put(0.91,-0.3){1} \put(1.91,-0.3){2} \
\put(2.91,-0.3){3} \put(3.91,-0.3){4} \put(4.91,-0.3){5} \
\put(5.91,-0.3){6} \put(6.91,-0.3){7} \put(7.91,-0.3){8} \
\put(8.91,-0.3){9} \put(9.91,-0.3){10}

\put(-0.25,-0.1){0} \put(-0.25,0.9){1} \put(-0.25,1.9){2} \
\put(-0.25,2.9){3} \put(-0.25,3.9){4} \put(-0.25,4.9){5} \
\put(-0.25,5.9){6} \put(-0.25,6.9){7} \put(-0.25,7.9){8} \
\put(-0.25,8.9){9} \put(-0.25,9.9){10}
}

\newcommand{\Fig}[3]{
\begin{figure}
\setlength{\unitlength}{1cm}
\centering{
\begin{picture}(8.5,5.5)
\thicklines
#1
\ifnum#3=1
\gridD
\fi
\end{picture}
}
\caption{#2}
\end{figure}
}

\begin{document}

\title{Non-iterative Type 4 and 5 Nonuniform FFT Methods in the One-Dimensional Case}

\author{J. Selva   
}

\maketitle

\markboth{}{}
\begin{abstract}

The so-called non-uniform FFT (NFFT) is a family of algorithms for efficiently computing
the Fourier transform of finite-length signals, whenever the time or frequency grid is
nonuniformly spaced. Following the usual classification, there exist five NFFT types.
Types 1 and 2 make it possible to pass from the time to the frequency domain with
nonuniform input and output grids respectively. Type 3 allows for both input and output
nonuniform grids. Finally, types 4 and 5 are the inverses of types 1 and 2 and are
expensive computationally, given that they involve iterative methods.  In this paper, we
solve this last drawback in the one-dimensional case by presenting non-iterative type 4
and 5 NFFT methods that just involve three NFFTs of types 1 or 2 plus some additional
FFTs. The methods are based on exploiting the structure of the Lagrange interpolation
formula. The paper includes several numerical examples in which the proposed methods are
compared with the Gaussian elimination (GE) and conjugate gradient (CG) methods, both in
terms of round-off error and computational burden.

\end{abstract}

\section{Introduction}

As is well known, the FFT computes the spectrum of a finite-length discrete signal with
complexity $\FO(P\log P)$, where $P$ is the signal's length. However, the FFT requires
regular grids in both the time (or spatial) and frequency domains. This constraint is a
shortcoming in various applications in which the sampling or evaluation grid is
nonuniform. The so-called nonuniform FFT (NFFT) is a family of algorithms for performing
the same transform as the FFT but with irregular input or output grids.  The NFFT has been
developed during the last decades for various applications
\cite{Anderson10,Kunis12,Zhou05}. There exist several surveys and tutorials on this topic
\cite{Keiner09,Potts03b,Benedetto03,Duijndam99}.

Following the classification in \cite{Dutt93c}, there are five basic types of NFFT. The
first three types have the same complexity order as the FFT and are non-iterative, while
types 4 and 5 require an iterative procedure like the conjugate gradient (CG) method,
\cite[Sec. 11.3]{Golub13}.  Type 1 performs the same operation as the FFT but with input
samples taken nonuniformly, while type 2 is similar to the inverse FFT but with nonuniform
output instants (or positions). Type 3 is a combination of types 1 and 2 and can be viewed
as a method to compute the spectrum of a nonuniform delta train at nonuniform
frequencies. Finally, types 4 and 5 are the inverses of types 1 and 2 respectively, and
are computed iteratively, usually through the CG method, thus being significantly more
expensive computationally than the first three types. Concretely, each iteration of the CG
method for either type 4 or 5 has the same complexity order as the FFT, but the number of
iterations is large. 

The purpose of this paper is to present two non-iterative methods for types 4 and 5 in the
one-dimensional case, which exploit the properties of the Lagrange interpolation
formula. They have the same complexity order as the FFT, $\FO(P\log P)$, and are far less
complex than the existing CG implementations. The extension of the proposed methods to
multiple dimensions seems difficult, given that there is no general Lagrange formula in
several variables \cite{Gasca01}.

The paper has been organized as follows. In the next section, we shortly recall the five
NFFT types, and introduce the Lagrange interpolation formula. Then, we present the
proposed NFFT methods in Secs. \ref{sec:elf} and \ref{sec:ctn}. These methods depend on
two parameters, the attenuation and oversampling factors, which are analyzed in
Sec. \ref{sec:sae}. Finally, Sec. \ref{sec:ne} contains a numerical assessment of the
methods' round-off error and complexity.

\subsection{Notation}

In the rest of the paper, we will use the following notation:

\begin{itemize}
\item Definitions will be introduced using the symbol ``$\equiv$''.
\item ``$\FO(\cdot)$'' will be the big-O notation.
\item $P$, $Q$, and $R$ will denote positive integers.
\item The notation $\SEQ{\cdot}pP$ will represent the vector formed by evaluating the
  expression inside curly braces for $p=0,\,,\ldots,\, P-1$. Thus, for a function
  $\Ff(x)$, $\SEQ{\Ff(p)}p P$ will be the vector $[\Ff(0),\,\Ff(1),\ldots,\, \Ff(P-1)]$.
\item The operators ``$\DFT$'' and ``$\IDFT$'' will respectively denote the DFT and IDFT of a
  vector. Thus, given a sequence $v_q$, $\DFT\SEQ{v_q}qP$ is the $P$-length vector whose
  element at position $p+1$ is
\be{eq:0}
\sum_{q=0}^{P-1}v_q\Fe^{-j2\pi pq/P}.
\ee
\item ``$\odot$'' will represent the element-by-element product of two vectors,
\be{eq:1}
\SEQ{v_p}pP\odot \SEQ{w_p}pP=\SEQ{v_pw_p}pP.
\ee

\item The operator ``$\mathrm{Coef}$'' will extract the coefficient vector of a given
  trigonometric polynomial,
\be{eq:2}
\mathrm{Coef}\sum_{p=0}^{P-1} F_p\pE{pt}=\SEQ{F_p}pP.
\ee

\item ``$\|\cdot\|$'' will refer to the quadratic norm,
\be{eq:3}
\big\|\SEQ{v_p}pP\big\|=\sqrt{\sum_{p=0}^{P-1}|v_p|^2.}
\ee

\item $\delta_p$ will be Kronecker's delta: $\delta_0=1$ and $\delta_p=0$ if $p\neq 0$.
  
\item The arrow ``$\rightarrow$'' will denote a replacement in a given expression.
\end{itemize}

\section{NFFT types and Lagrange interpolation formula}

The NFFT methods perform the same operations as the FFT or IFFT but allow for nonuniform
sampling or evaluation grids. Basically, there exist five NFFT types, that depend on
which grid is nonuniform. Following the classification in \cite{Dutt93c}, the type-1 NFFT
transforms a nonuniform into a uniform grid, and can be viewed as a method to sample
regularly the spectrum of a nonuniform delta train like
\be{eq:33}
\Falp(t)\equiv \sum_{p=1}^{Q} a_p\Fdel(t-t_p),
\ee
where $\SEQ{a_{p+1}}pQ$ and $\SEQ{t_{p+1}}pQ$ are a complex and real vector respectively,
with ${0\leq t_p<1}$.  More precisely, if $\FA(f)$ denotes the
spectrum of $\Falp(t)$,
\be{eq:32}
\FA(f) \equiv \sum_{p=1}^Q a_p\mE{ft_p},
\ee
the type-1 NFFT computes $\SEQ{\FA(p)}pP$ from $\SEQ{t_{p+1}}pQ$ and $\SEQ{a_{p+1}}pQ$
with complexity $\FO(P\log P+Q)$. Typically, it involves one gridding operation and one
FFT, both with oversampling factor two. Fundamentally, the gridding operation consists of
replacing the deltas in (\ref{eq:33}) with band-limited discrete pulses.

One of the basic applications of the type-1 NFFT is to evaluate nonuniform
convolutions of the form
\bae{eq:26}{1.5}{r@{\,}l}
\D
\Fgam(t)&\D\equiv \sum_{p=1}^Q a_p\Flam(t-t_p)
\eae
in a regular grid $\SEQ{q/P}qP$, where $\Flam(t)$ is a trigonometric polynomial
\be{eq:4}
\Flam(t)\equiv \sum_{p=0}^{R-1}\Lambda_p\pE{pt},
\ee
and $R$ is an integer multiple of $P$, ($R\equiv \eta P$ with integer $\eta\geq 1$). To
see the relation between $\Fgam(t)$ and the type-1 NFFT, note that $\Fgam(t)$ is the
convolution of $\Flam(t)$ with the delta train in the type-1 NFFT in (\ref{eq:33}),
\be{eq:5}
\Fgam(t)=\Flam(t)*\sum_{p=1}^Q a_p\Fdel(t-t_p)=\Flam(t)*\Falp(t),
\ee
Thus, we have that $\Fgam(t)$ is the polynomial 
\be{eq:6}
\Fgam(t)\equiv \sum_{p=0}^{R-1}\Lambda_p\FA(p)\pE{pt},
\ee
in which $\SEQ{\FA(p)}pR$ can be obtained by means of one type-1 NFFT with $P\rightarrow
R$. So, we have from (\ref{eq:6}) that $\SEQ{\Fgam(q/P)}qP$ is the output of one inverse
DFT,
\be{eq:7}
\SEQ{\Fgam(q/P)} q P=P\, \IDFT \Big\{\sum_{r=0}^{\eta-1}
\Lambda_{Pr+p}B_{Pr+p}\Big\}_p^P.
\ee

The type-1 NFFT is a basic tool in spectral estimation from nonuniform samples
\cite{Babu10} and, additionally, has various application fields like synthetic aperture
radar (SAR) imaging \cite{Anderson10}, graphics processing \cite{Kunis12}, and magnetic
resonance imaging (MRI) \cite{Fessler03}.

The type-2 NFFT is complementary to the type-1 NFFT in the sense that it converts a
uniform into a nonuniform grid. Specifically, given a trigonometric polynomial of the form
\be{eq:34}
\Fs(t)\equiv \sum_{p=0}^{P-1}S_p \Fe^{j2\pi p t},
\ee
where $\SEQ{S_p}pP$ is a complex vector, the type-2 NFFT computes $\Fs(t)$ at $Q$
arbitrary instants $\SEQ{t_{p+1}}pQ$ with complexity $\FO(P\log P+Q)$. Usually, it
involves one weighting of the vector $\Sp$, followed by one inverse FFT, and one final
interpolation operation. This last operation consists of a short summation extended to the
samples close to the interpolation instants $t_{p+1}$, and all the processing is performed
with oversampling factor two.  As (\ref{eq:34}) shows, its basic use is signal
interpolation from spectral samples. It has applications in antenna design
\cite{Capozzoli14}, computational electromagnetics \cite{Liu98b}, array processing
\cite{Selva05a}, among other fields.

The type-3 NFFT is a combination of the previous two types and is a nonuniform to
nonuniform transformation.  In short, given the instants $\SEQ{t_{p+1}}pQ$ and coefficients
$\SEQ{a_{p+1}}pQ$ in (\ref{eq:33}), the type-3 NFFT computes $\SEQ{\FA(f_{r+1})}rR$ for a
given set of frequencies $\SEQ{f_{r+1}}rR$ with complexity $\FO(P\log P+Q+R)$. It is
applied in heat flow computation and MRI \cite{Lee05}.

Finally, types 4 and 5 are the inverses of types 1 and 2 respectively, assuming
${P=Q}$, and are usually computed iteratively by means of the conjugate gradient (CG)
method. Specifically, from (\ref{eq:32}) the type-4 NFFT inverts the linear system
\be{eq:47}
\FA(p) = \sum_{q=1}^P a_q\mE{pt_q},\;p=1,\,\ldots,\,P,
\ee
where the unknowns are $\SEQ{a_{q+1}}qP$.  The existing CG method for this NFFT type
locates the solution of (\ref{eq:47}) iteratively and in each iteration requires a
discrete convolution with complexity $\FO(P\log P)$. The number of iterations required is
usually large.
From (\ref{eq:34}), the type-5 NFFT inverts the system
\be{eq:48}
\Fs(t_q)= \sum_{p=0}^{P-1}S_p \Fe^{j2\pi p t_q},\;q=0,\,1,\ldots,\,P-1,
\ee
and the existing CG method's performance is similar to that for type 4. One basic use of
this NFFT type is to re-sample a signal to a regular grid \cite{Selva15}. Types 4 and 5
appear in computational electromagnetics \cite{Zhou05}, MRI \cite{Fessler03}, and spectral
estimation \cite{Babu10}, among other fields. 

The purpose of this paper is to present two non-iterative methods for the type-4 and
type-5 NFFTs in the one-dimensional case, which are significantly less expensive
computationally than the state-of-the-art methods. They are based on the properties of the
Lagrange formula
\be{eq:9}
\Fs(t)=\sum_{p=1}^P\Fs(t_p)\frac{\FL(\Fe^{j2\pi t})}{\FL'(\Fe^{j2\pi t_p})(\Fe^{j2\pi
    t}-\Fe^{j2\pi t_p})},
\ee
where $\FL(\pE t)$ is the kernel
\be{eq:11}
\FL(\pE t)\equiv \prod_{p=1}^P\pE t-\pE{t_p}.
\ee
Fundamentally, they consist of writing (\ref{eq:9}) in terms of nonuniform convolutions
and interpolation operations, which can be efficiently computed through type 1 and 2
NFFTs.

\section{Computation of the type-5 NFFT}
\label{sec:elf}

\begin{figure}
  
\setlength{\unitlength}{8.5cm}

\centering{
%\fbox{
\begin{picture}(0.9,1.24)
\thicklines

\put(0.39,1.21){\fbox{$\D \SEQ{t_{p+1}} p P$}}

\put(0.47,1.17){\vector(0,-1){0.1}}
\put(0.5,1.12){type-1 NFFT, (\ref{eq:28})}

\put(0.21,1){$\D\big\{\sum_{p=1}^P \log(1-\pE{(q/P-t_p+ja)})\big\}_q^P$}

\put(0.47,0.96){\vector(0,-1){0.1}}
\put(0.5,0.9){(\ref{eq:29})}

\put(0.32,0.8){$\D\LK$}
\put(0.47,0.77){\vector(0,-1){0.1}}
\put(0.5,0.71){(\ref{eq:43}), type-2 NFFT}

\put(0.31,0.81){\line(-1,0){0.2}}
\put(0.31,0.81){\vector(-1,0){0.1}}
\put(0.11,0.8116){\line(0,-1){0.603}}
\put(0.11,0.81){\vector(0,-1){0.32}}
\put(0.11,0.21){\vector(1,0){0.21}}

\put(0.02,0.5){(\ref{eq:30})}

\put(0.37,0.62){$\D\Ld$}
\put(0.47,0.59){\vector(0,-1){0.1}}
\put(0.5,0.53){(\ref{eq:27}), type-1 NFFT}

\put(0.12,0.41){$\LSb$}

\put(0.47,0.35){\vector(0,-1){0.1}}
\put(0.5,0.29){(\ref{eq:30})}

\put(0.33,0.2){$\D\sqa$}
\put(0.47,0.17){\vector(0,-1){0.1}}
\put(0.5,0.11){(\ref{eq:31})}

\put(0.42,0.02){$\Sp$}

\put(0.14,0.68){\fbox{$\SEQ{\Fs(t_{p+1}}pP$}}
\put(0.24,0.64){\vector(0,-1){0.14}}

\end{picture}
%}
}

\caption{\label{fig:d5} Flow diagram of type-5 NFFT method.}
\end{figure}
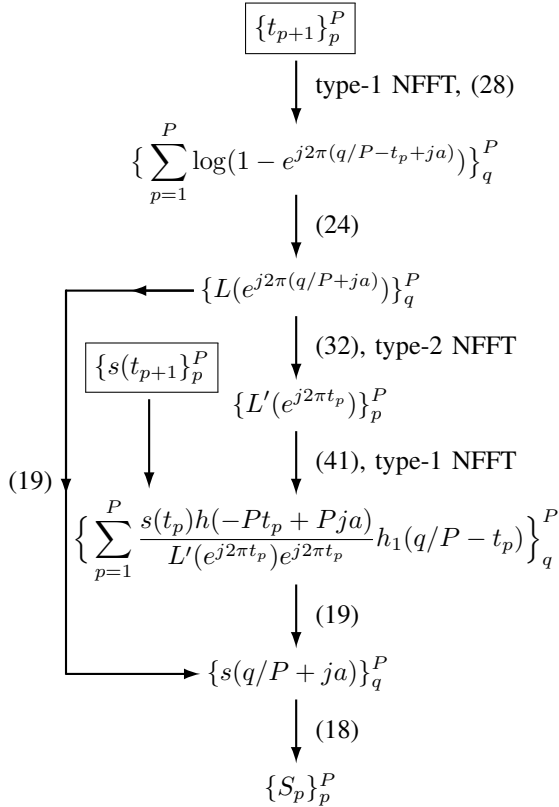

In this section, we derive the proposed method for the type-5 NFFT, and in the next
section the corresponding method for the type 4. We follow this order for simplicity,
given that the latter is a simple corollary of the former. Since the derivations in this
section are quite detailed, they have been split into separate sub-sections and in each of
them one of the transitions in Fig. \ref{fig:d5} is proved, starting with the last one and
proceeding backward. Note that the inputs in that diagram (framed vectors) are the type-5
NFFT inputs, and the final output is the required vector of coefficients $\SEQ{S_p}pP$.

Basically, the proposed type-5 NFFT method can be regarded as a procedure to evaluate the
Lagrange formula in (\ref{eq:9}) simultaneously at all the instants in a regular grid
$\SEQ{q/P+ja}pP$, where $ja$ is an imaginary shift, ${a>0}$. This simultaneous evaluation
is performed through convolutions like (\ref{eq:26}), type-2 NFFTs, DFTs, and other
simpler operations. In the next sub-section, we start by relating the type-5 NFFT output
$\Sp$ with the sequence $\sqa$, which is then computed in the sub-sections that follow.

\subsection{Coefficients $S_p$ from samples $\Fs(q/P+ja)$}

The final output $\Sp$ can be computed from the sequence ${\sqa}$ through one DFT, where
$ja$ is a fixed imaginary shift, $a>0$.  This is so because $\Fs(t+ja)$ is just the output
of passing $\Fs(t)$ through an invertible filter, as the following equation reveals
\be{eq:8}
\Fs(t+ja)=\sum_{p=0}^{P-1}S_p\Fe^{-2\pi p a}\pE{pt}.
\ee
The coefficients $S_p$ are given by
\be{eq:31}
\SEQ{S_p}p P = \frac{1}{P}\SEQ{\Fe^{2\pi p a}} p P\odot \DFT  \SEQ{\Fs(q/P+ja)} q P,
\ee
and the problem comes down to computing ${\sqa}$.  As will be shown in the sequel, the
shift $ja$ is key in the method, given that it eliminates the singularities in two
convolution kernels.

\subsection{Sequence $\Fs(q/P+ja)$ from Lagrange formula factors}
\label{sec:ssl}

In order to compute ${\sqa}$, let us start by writing the Lagrange formula
in (\ref{eq:9}) for $t=q/P+ja$ in the following way, 
\begin{multline}
\label{eq:30}
  \Fs(q/P+ja)=\FL(\Fe^{j2\pi (q/P+ja)})\\
\cdot\sum_{p=1}^P\frac{\Fs(t_p)}{\FL'(\Fe^{j2\pi t_p})(\Fe^{j2\pi
    (q/P+ja)}-\Fe^{j2\pi t_p})}.
\end{multline}
Note that this is the product of two terms and none of them has any poles or zeros for
real $t$ due to the shift $ja$. Thus, it is possible to compute $\SEQ{\Fs(q/P+ja)} q P$ by
multiplying element-wise the sequences
\be{eq:10}
\LK
\ee
and
\be{eq:24}
\LS.
\ee

In the next sub-section, we show how to compute (\ref{eq:10}). Then, in Sec.  \ref{sec:ds}
we address the computation of the samples $\Ld$ appearing in  (\ref{eq:24}),
and in sub-section \ref{sec:s} the computation of the whole sequence (\ref{eq:24}).

\subsection{Kernel samples $\FL(\Fe^{j2\pi (q/P+ja)})$ from a nonuniform convolution.}
\label{sec:ks}

Let us analyze the kernel $\FL(\pE{(t+ja)})$ in order to compute (\ref{eq:10}).  From
(\ref{eq:11}), we have that the logarithm of $\FL(\pE{(t+ja)})$ is the following:
\bae{eq:12}{1.5}{l}
\D\log \FL(\pE{(t+ja)})=\D\sum_{p=1}^P\log(\pE{(t+ja)}-\pE{t_p})\\
{}\hspace{0.5cm}=\D jP\pi+j2\pi\sum_{p=1}^Pt_p+\sum_{p=1}^P \log(1-\pE{(t-t_p+ja)}).
\eae
Let $\Fv(t)$ denote the last summation
\be{eq:13}
\Fv(t)\equiv \sum_{p=1}^P \log(1-\pE{(t-t_p+ja)}).
\ee
From the last two equations, we have that the kernel samples can be computed from the
sequence $\vseq$ through the equation
\be{eq:29}
\FL(\Fe^{j2\pi (q/P+ja)})=\exp
\Big(jP\pi+j2\pi\sum_{p=1}^Pt_p+\Fv(q/P)\Big).
\ee

\subsection{Computation of $\Fv(q/P)$.}
\label{sec:cv}

Let us analyze the definition of $\Fv(t)$ in (\ref{eq:13}) in order to compute
$\SEQ{\Fv(q/P)}qP$. The summation in that definition can be written as the convolution of
a delta train with the kernel $\log(1-\pE{(t+ja)})$,
\begin{align}
  \Fv(t) &=\log(1-\pE{(t+ja)})*\sum_{p=1}^P \Fdel(t-t_p)\nonumber\\
\label{eq:19}  &=\log(1-\pE{(t+ja)})*\Fz(t), 
\end{align}
where
\be{eq:14}
\Fz(t)\equiv \sump \Fdel(t-t_p).
\ee
Besides, the coefficients of
$\log(1-\pE{(t+ja)})$ decay with exponential trend as its Fourier series reveals,
\be{eq:42}
\log(1-\pE{(t+ja)})=-\sum_{p=1}^\infty \frac{1}{p}\Fe^{-2\pi p a}\pE{pt}.
\ee
Therefore, in (\ref{eq:19}) we may replace $\log(1-\pE{(t+ja)})$ with this last series,
but truncated at an index ensuring a negligible approximation error. Specifically, we have
the approximation
\be{eq:28}
\Fv(t)\approx\Fg(t)*\Fz(t),
\ee
where
\be{eq:41}
\Fg(t)\equiv -\sum_{p=1}^{R-1} \frac{1}{p}\Fe^{-2\pi p a}\pE{pt}
\ee
and $R$ is the index of the first neglected coefficient. For simplicity, we take $R$ equal
to an integer multiple of $P$, $R=\eta P$, (integer $\eta\geq 1$). (\ref{eq:28}) is a
nonuniform convolution of the form in (\ref{eq:26}). Thus, we may compute
$\SEQ{\Fv(q/P)}qP$ using the procedure already described for (\ref{eq:26}) with $\Flam(t)
\rightarrow \Fg(t)$, $a_p\rightarrow 1$, and $R\rightarrow\eta P$.

\subsection{Derivative samples $\FL'(\pE{t_p})$ from kernel samples $\FL($ $\pE{(q/P+ja)})$}
\label{sec:ds}

Let $\SEQ{L_p} p {P+1}$ denote the set of coefficients of $\FL(\pE t)$ and note the following
straight-forward equations
\begin{gather}
\label{eq:16}
\mathrm{Coef}\, \FL'(\pE t)=\SEQ{(p+1)L_{p+1}} p P\\
\label{eq:15}
\mathrm{Coef}\, \FL(\pE {(t+ja)})=\SEQ{\Fe^{-2\pi p a}L_p} p {P+1}. 
\end{gather}
The last equation implies that the DFT of $\SEQ{\FL(\pE{(q/P+ja)})} q P$ gives the
coefficients $\SEQ{L_p} p P$ with a weighting. More precisely, we have to take the DFT of
$\SEQ{\FL(\pE{(q/P+ja)})} q P$, remove the aliasing at $p=0$, and compensate the factor
$\Fe^{-2\pi p a}$ in (\ref{eq:15}). The result is the following equation
\begin{multline}
  \label{eq:43}
  \SEQ{L_p} p P =\frac{1}{P}\Big(\DFT \SEQ{\FL(\pE{(q/P+ja)})} q P
  -P\Fe^{-2\pi Pa}\SEQ{\Fdel_p} p P\Big)\\
  \odot
\SEQ{\Fe^{2\pi pa}} p P.
\end{multline}
Once the coefficients $\SEQ{L_p} p P$ are known, and noting that $L_P=1$, we may obtain
$\Ld$ by applying a type-2 NFFT to $\SEQ{(p+1)L_{p+1}} p P$, due to (\ref{eq:16}).

\subsection{Summation in (\ref{eq:24}) assuming known samples $\FL'(\pE{t_p})$}
\label{sec:s}

In order to compute the summation in (\ref{eq:24}), let us first define the kernel 
\be{eq:17}
\Fh(t)\equiv\frac{1}{\pE{t}-1},
\ee
and re-write the summation in that equation in the following way
\begin{gather}
\label{eq:21}
  \sum_{p=1}^P\frac{\Fs(t_p)}{\FL'(\Fe^{j2\pi
    t_p})\pE{t_p}}\Fh(q/P-t_p+ja)\\
\label{eq:18}
{}\hfill =\Fh(t+ja)*\sum_{p=1}^P\frac{\Fs(t_p)}{\FL'(\Fe^{j2\pi
    t_p})\pE{t_p}}\Fdel(q/P-t_p).
\end{gather}
Note that this expression resembles that in Sec. \ref{sec:cv}, Eq. (\ref{eq:19}), for the
computation of the samples $\vseq$. In it, we have the convolution of a signal $\Fh(t+ja)$
with a nonuniform delta train. Besides, $\Fh(t+ja)$ has infinite bandwidth, as can be
readily seen in its Fourier series
\be{eq:20}
\Fh(t+ja)=-\sum_{r=0}^\infty \Fe^{-2\pi r a}\pE{rt},
\ee
and the delta train coefficients are known, given that $\Ld$ has been pre-computed in
Sec. \ref{sec:ds}. However, (\ref{eq:21}) is a simpler case because $\Fh(t+ja)$ can be
replaced by a band-limited kernel, and this allows us to employ a type-1 NFFT without any
truncation. To see this point, let us insert a term $\Fh(q-Pt_p+Pja)$ in the summand in
(\ref{eq:21}) and operate as follows:
\bae{eq:22}{1.9}{l}
\D\frac{\Fs(t_p)}{\FL'(\Fe^{j2\pi
    t_p})\pE{t_p}}\Fh(q/P-t_p+ja)\hspace{2cm}{}\\
{}\hfill\D =\frac{\Fs(t_p)\Fh(q-Pt_p+Pja)}{\FL'(\Fe^{j2\pi
    t_p})\pE{t_p}}\cdot\frac{\Fh(q/P-t_p+ja)}{\Fh(q-Pt_p+Pja)}.\\
\eae
Note that in the first fraction we may simplify
\be{eq:23}
\Fh(q-Pt_p+Pja)=\Fh(-Pt_p+Pja),
\ee
and the second fraction is equal to $\Fh_1(q/P-t_p)$, where $\Fh_1(t)$ is a new
band-limited pulse, defined by
\bae{eq:36}{1.5}{r@{\,}l}
\D\Fh_1(t)&\equiv \D\frac{\Fh(t+ja)}{\Fh(P(t+ja))}=\frac{\pE{P(t+ja)}-1}{\pE{(t+ja)}-1}\\
&\D=\sum_{p=0}^{P-1}\Fe^{-2\pi p a}\pE{p
  t}.
\eae
So, we have the following formula for the vector in (\ref{eq:24})
\bae{eq:25}{1.5}{l}
\D\LS  \\
\D=
\Big\{\sum_{p=1}^P\frac{\Fs(t_p)\Fh(-Pt_p+Pja)}{\FL'(\Fe^{j2\pi
t_p})\pE{t_p}}\Fh_1(q/P-t_p)\Big\}_q^P.
\eae
This is a non-uniform convolution that can be computed through a type-1 NFFT, if we
identify in (\ref{eq:26})
\bae{eq:27}{1.5}{rll}
\D R&\D\rightarrow & P\\
\D a_p&\D\rightarrow& \D\frac{\Fs(t_p)\Fh(-Pt_p+Pja)}{\FL'(\Fe^{j2\pi
t_p})\pE{t_p}}\\
\D\Flam(t)&\D\rightarrow & \D\Fh_1(t).
\eae

\begin{figure}
  
\setlength{\unitlength}{8.5cm}
\centering{
%\fbox{
\begin{picture}(0.9,1.58)
\thicklines

\put(0.235,1.55){\fbox{$\D \SEQ{t_{p+1}} p P$}}

\put(0.32,1.51){\vector(0,-1){0.1}}
\put(0.35,1.46){type-1 NFFT, (\ref{eq:28})}

\put(0.06,1.34){$\D\big\{\sum_{p=1}^P \log(1-\pE{(q/P-t_p+ja)})\big\}_q^P$}

\put(0.32,1.3){\vector(0,-1){0.1}}
\put(0.35,1.24){(\ref{eq:29})}

\put(0.17,1.14){$\D\LK$}
\put(0.32,1.12){\vector(0,-1){0.1}}
\put(0.35,1.06){(\ref{eq:43}), type-2 NFFT}

\put(0.16,1.15){\line(-1,0){0.15}}
\put(0.16,1.15){\vector(-1,0){0.09}}
\put(0.01,1.1516){\line(0,-1){0.603}}
\put(0.01,1.15){\vector(0,-1){0.32}}
\put(0.01,0.55){\vector(1,0){0.18}}

\put(0.02,0.84){(\ref{eq:30})}

\put(0.21,0.97){$\D\Ld$}

\put(0.12,0.75){$\Big\{\D\sum_{p=1}^Pa_p\Fh_1(q/P-t_p)\Big\}_q^P$}
\put(0.65,0.76){\vector(-1,0){0.1}}
\put(0.67,0.75){\fbox{$\SEQ{\FA(p)}pP$}}

\put(0.57,0.71){(\ref{eq:37})}

\put(0.32,0.69){\vector(0,-1){0.1}}
\put(0.35,0.63){(\ref{eq:30})}

\put(0.28,0.36){$\Sp$}
\put(0.32,0.51){\vector(0,-1){0.1}}
\put(0.35,0.45){(\ref{eq:31})}

\put(0.2,0.54){$\D\sqa$}

\put(0.32,0.32){\vector(0,-1){0.1}}
\put(0.35,0.27){type-2 NFFT}

\put(0.255,0.17){$\SEQ{\Fs(t_p)}pP$}

\put(0.32,0.14){\vector(0,-1){0.1}}
\put(0.275,0){$\SEQ{a_p}pP$}

\put(0.47,0.98){\line(1,0){0.4}}
\put(0.47,0.98){\vector(1,0){0.23}}
\put(0.87,0.9816){\line(0,-1){0.973}}
\put(0.87,0.9816){\vector(0,-1){0.4865}}
\put(0.87,0.01){\vector(-1,0){0.48}}

\put(0.78,0.505){(\ref{eq:40})}
\put(0.34,0.09){(\ref{eq:40})}

\end{picture}
%}
}

\caption{\label{fig:d4} Flow diagram of type-4 NFFT method.}
\end{figure}
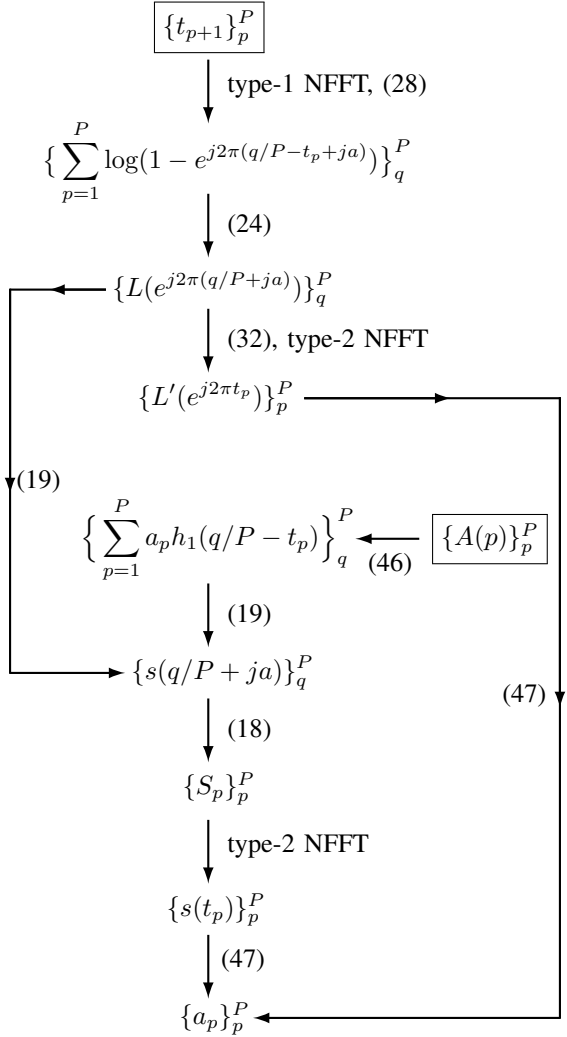

\section{Computation of the type-4 NFFT}
\label{sec:ctn}

In the type-4 NFFT, we set $Q=P$ in (\ref{eq:33}) and (\ref{eq:32}) and the objective is
to compute $\SEQ{a_{q+1}}qP$ given $\SEQ{A(p)}pP$. In order to derive the type-4 NFFT
method, note first that there is a convolution similar to (\ref{eq:26}) in the derivation
of the type-5 NFFT, specifically, in the computation of the sequence
\be{eq:35}
\LSb;
\ee
(see Fig. \ref{fig:d5}). Let us relate this sequence with the known spectral samples
$\SEQ{A(p)}pP$. For this, consider the unique signal $\Fs(t)$ of the form in (\ref{eq:34})
such that
\be{eq:39}
\SEQ{\Fs(t_{p+1})}pP=\Big\{\frac{a_{p+1} \FL'(\pE{t_{p+1}})\pE{t_{p+1}}}{\Fh(-Pt_{p+1}+Pja)}\Big\}_p^P,
\ee
where $\SEQ{a_{p+1}}pP$ is the unknown sequence. For this signal, (\ref{eq:35})
takes the form
\be{eq:38}
\Big\{\D\sum_{p=1}^Pa_p\Fh_1(q/P-t_p)\Big\}_q^P.
\ee
Let us apply the DFT to this sequence. Recalling (\ref{eq:36}), we have 
\begin{multline}
\DFT 
  \Big\{\D\sum_{p=1}^Pa_p\Fh_1(q/P-t_p)\Big\}_q^P\\
  =\sum_{p=1}^Pa_p\DFT\SEQ{\Fh_1(q/P-t_p)}qP\\
  =\sum_{p=1}^Pa_p\SEQ{P\Fe^{-2\pi r a}\Fe^{-j2\pi t_pr}}rP\\
  =P\Big\{\Fe^{-2\pi r a}\sum_{p=1}^Pa_p\Fe^{-j2\pi t_pr}\Big\}_r^P
  =P\SEQ{\Fe^{-2\pi r a}\FA(r)}rP\\
  =P\SEQ{\Fe^{-2\pi r a}}rP \odot \SEQ{\FA(r)}rP.
\end{multline}

Thus, applying the IDFT, we have
\begin{multline}
\label{eq:37}
\Big\{\D\sum_{p=1}^Pa_p\Fh_1(q/P-t_p)\Big\}_q^P \\ =P\,\IDFT\Big(\SEQ{\Fe^{-2\pi r a}}rP \odot \SEQ{\FA(r)}rP\Big).
\end{multline}
Using this equation, we may first compute (\ref{eq:38}) from $\SEQ{\FA(r)}rP$ without
actually knowing $\SEQ{a_{p+1}}pP$. Then, we may follow the diagram in Fig. \ref{fig:d5},
successively computing $\sqa$ and $\Sp$. Once $\Sp$ is available, we may compute
$\SEQ{\Fs(t_{p+1})}pP$ through a type-2 NFFT. And finally, we may obtain $\SEQ{a_{p+1}}pP$
by inverting (\ref{eq:39}),
\be{eq:40}
\SEQ{a_{p+1}}pP=\Big\{\frac{\Fs(t_{p+1})\Fh(-Pt_{p+1}+Pja)}
{\FL'(\pE{t_{p+1}})\pE{t_{p+1}}}\Big\}_p^P.
\ee
Fig \ref{fig:d4} shows the flow diagram of this type-4 NFFT method.

\section{Selection of $a$ and $\eta$ and refined methods}
\label{sec:sae}

The methods in the previous two sections deliver the type 4 and 5 NFFTs with an accuracy
only limited by the truncation of (\ref{eq:42}), assuming infinite working
precision. However, for finite precision an incorrect selection of the damping factor $a$
and oversampling factor $\eta$ may completely spoil the final result. We can see this
point by analyzing the computation of $\SEQ{\Fv(q/P)}qP$ in Sec. \ref{sec:cv}. On the one
hand, the truncation of (\ref{eq:42}) requires a negligible ratio between the first and
last summand of (\ref{eq:41}). Thus, if $\mu$ denotes this last ratio, defined by
\be{eq:49}
\mu\equiv \frac{\Fe^{-2\pi (\eta P-1)a}}{\eta
P-1},
\ee
then $a$ and $\eta$ should selected to ensure $\mu$ is close to the working precision. So,
we may see from (\ref{eq:49}) that, in rough terms, the product $aP\eta$ must be
sufficiently large, and this can be achieved by either increasing $a$ or $\eta$. However,
an increase in $a$ produces a strong attenuation in the computation of the coefficients
$\SEQ{L_p}pP$ in (\ref{eq:43}) due to the vector $\SEQ{\Fe^{2\pi pa}} p P$. And an
increase in $\eta$ seems suitable for truncating  (\ref{eq:42}) while reducing the
damping effect in (\ref{eq:43}), but there is a corresponding increase in the
computational burden, because the computation of $\SEQ{\Fv(q/P)}qP$ involves a type-1 NFFT
of size $\eta P$.

A simple way to overcome this situation consists of selecting $a$ and $\mu$ that produce
an inaccurate result, and then applying the method twice, first to the input sequences
$\SEQ{t_{p+1}}pP$,\,$\SEQ{a_{p+1}}pP$, and then to $\SEQ{t_{p+1}}pP$ and the residual
error, which can be computed through a type-1 or type-2 NFFT. More precisely, suppose we
require to compute the type-4 NFFT, but the method available produces an error
$\SEQ{\Phi_{0,p}}pP$,
\be{eq:44}
\SEQ{t_{p+1}}pP,\,\SEQ{a_{p+1}}pP\overset{\text{type-4 method}}{\xrightarrow{\hspace*{2cm}}} \SEQ{\FA(p)+\Phi_{0,p}}pP.
\ee
Also, suppose that the accuracy of this method is $\epsilon<1$ in the sense that 
\be{eq:45}
\big\|\SEQ{\Phi_{0,p}}pP\big\|<\epsilon\,\big\|\SEQ{\FA(p)}pP\big\|
\ee
for any possible input sequence $\SEQ{a_{p+1}}pP$.  Then, we may obtain a more accurate
approximation of $\SEQ{\FA(p)}pP$ in the following steps:

\begin{enumerate}

\item Compute the type-1 NFFT of the right-hand side of (\ref{eq:44}), 
\[
\hspace{-1cm}\SEQ{t_{p+1}}pP,\;
\SEQ{\FA(p)+\Phi_{0,p}}pP\overset{\text{type-1 NFFT}}{\xrightarrow{\hspace*{1.5cm}}}
\SEQ{a_{p+1}+\phi_{0,p+1}}pP,
\]
where $\SEQ{\phi_{0,p+1}}pP$ is the type-1 NFFT of $\SEQ{t_{p+1}}pP$ and
$\SEQ{\Phi_{0,p}}pP$.

\item Subtract $\SEQ{a_{p+1}}pP$ to the last output to obtain $\SEQ{\phi_{0,p+1}}pP$.
\item Apply the type-4 method to $\SEQ{\phi_{0,p+1}}pP$,

\[
\SEQ{\phi_{0,p+1}}pP\overset{\text{type-4 method}}{\xrightarrow{\hspace*{2cm}}}
\SEQ{\Phi_{0,p}+\Phi_{1,p}}pP
\]

where $\SEQ{\Phi_{1,p}}pP$ is a new residual error. Now we have
\be{eq:46}
\big\|\SEQ{\Phi_{1,p}}pP\big\|<\epsilon^2\,\big\|\SEQ{\FA(p)}pP\|.
\ee
\item Subtract the last sequence from the output of (\ref{eq:45}). The result is
  $\SEQ{\FA(p)-\Phi_{1,p}}pP$ and we have doubled the accuracy due to (\ref{eq:46}).
  
\end{enumerate}

A similar refinement can be applied to the type-5 NFFT. 

\section{Numerical examples}
\label{sec:ne}

In this section, we evaluate the proposed NFFT methods in terms of round-off error and
computational burden. We only present results for the type-4 NFFT, though they also
represent the corresponding performance of the type-5 NFFT. Actually, the figures that
follow and those for the type-5 NFFT are identical and, therefore, it is redundant to
repeat them in the present paper.  This identical performance is due to the relation
between the linear systems in (\ref{eq:47}) and (\ref{eq:48}. As can be readily inferred
from these last equations, the linear system solved by the types 4 and 5 NFFT form a dual
pair, i.e, if we take the Hermitian of the type-4 linear system matrix we obtain the
corresponding type-5 matrix. A consequence of this duality is that methods like Gaussian
elimination and conjugate gradient have identical round-off error performance and
computational burden for both types. And this is also true for the NFFT methods proposed
in this paper. As can be easily deduced from Figs. \ref{fig:d5} and \ref{fig:d4}, the
type-4 and type-5 evaluation procedures are the same except for a small modification that
involves no change in either the round-off error performance or computational burden.

We have evaluated the performance of the type-4 (and type-5) NFFT method in the following
setup: 

\begin{itemize}

\item \emph{Monte Carlo trials.} The figures have been generated from just 10 Monte Carlo
  trial, given that the large values of $P$ produce low-variance round-off error
  estimates. In these trials, the sampling instants $\SEQ{t_{p+1}}pP$ where obtained by
  shifting the elements of a regular grid with spacing $1/P$. These shifts were
  independent and had uniform distribution in the interval $[0,0.6/P]$. The amplitudes
  $\SEQ{a_{p+1}}pP$ were independent complex Gaussian samples of zero mean and variance
  one.

\item \emph{Numerical methods.} We have used the following methods:

\begin{itemize}
\item GE: Computation based on inverting the associated linear system through Gaussian
  elimination, \cite[Ch. 3]{Golub13}.
\item CG: The same inversion but using the conjugate gradient method \cite[p. 73]{Kunis06b}. 
\item NFFT: Method proposed in this paper in Sec. \ref{sec:ctn}.
\item R-NFFT: Previous method with refinement (Sec. \ref{sec:sae}).
\end{itemize}

\item \emph{Computational burden.} We have measured the computational burden in
  floating-point operations (flops), following the counts in Fig. \ref{fig:fc} for basic
  operations.

\item \emph{Round-off error measure.} Given a true vector $\SEQ{\FA(p)}pP$ and an
    interpolated vector $\SEQ{\tilde A_p}pP$, the error measure has been
\[
\frac{\|\SEQ{\FA(p)-\tilde A_p}pP\|}{\|\SEQ{\FA(p)}pP\|}.
\]

\begin{figure}
\centering{ \begin{tabular}{l|c}
Operation & Flops\\\hline
Real sum & 1\\
Complex sum & 2\\
Real multiplication & 1\\
Complex multiplication & 6\\
Complex exponential & 7\\
Size-$N$ FFT, IFFT & $5N\log_2N$\\
\end{tabular}}
\caption{\label{fig:fc} Flop counts for basic operations.}
\end{figure}

\end{itemize}

\Fig{
\put(0,0){\includegraphics{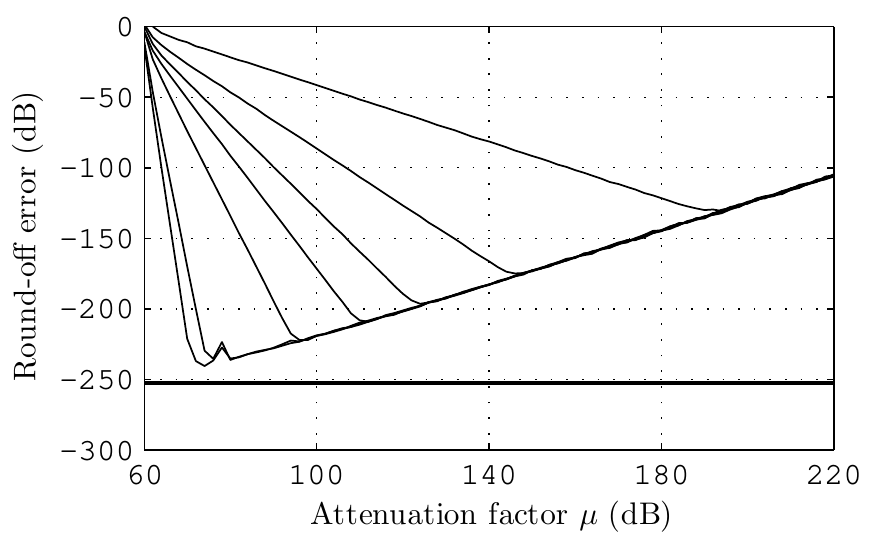}}
\put(5.15,4.1){$\eta=1$}

\put(4.15,3.45){$\eta=2$}
\put(3.75,2.91){$3$}
\put(3.35,2.65){$4$}

\put(2.9,2.3){$6$}
\put(2,2.2){$15$}
\put(1.5,1.8){$20$}

\put(5.5,1.75){GE, CG methods}
}{
\label{fig:1} Round-off error versus attenuation factor ($\mu$) for several 
oversampling factors ($\eta$) for the type-4 NFFT method.
}{0}
Fig. \ref{fig:1} shows the round-off error of the NFFT method versus the attenuation $\mu$
in (\ref{eq:49}) for several oversampling factors $\eta$ and $P=1024$. This figure also
includes the round-off error of the GE and CG methods as benchmarks. Note that with
$\eta=1$, the achievable round-off error is around $-130$ dB, which is a value
significantly larger that the GE, CG benchmarks. However, this error decreases with
$\eta$. With $\eta=6$ it is around -220 dB and can be reduced by increasing $\eta$ to a
value only 10 dB above the previous benchmarks.
\Fig{
\put(0,0){\includegraphics{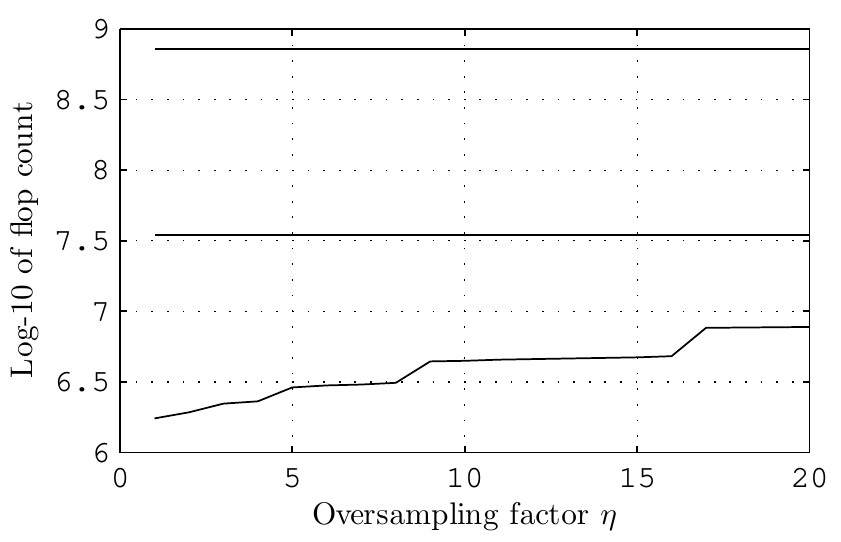}}
\put(2,4.7){GE method}
\put(2,2.8){CG method}
\put(5,1.5){type-4 NFFT}
}{
\label{fig:7} Computational burden of type-4 NFFT method, measured in floating
point operations (flops), versus oversampling factor $\eta$.  
5  }{0}
As Fig. \ref{fig:7} shows, this decrease is obtained at the expense
of a higher computational burden but, by far, the type-4 NFFT is the cheapest
computationally. Actually, its computational burden is more than factor 10 smaller that
the CG method's complexity for $\eta=6$.

\begin{figure}
  \centering{\includegraphics{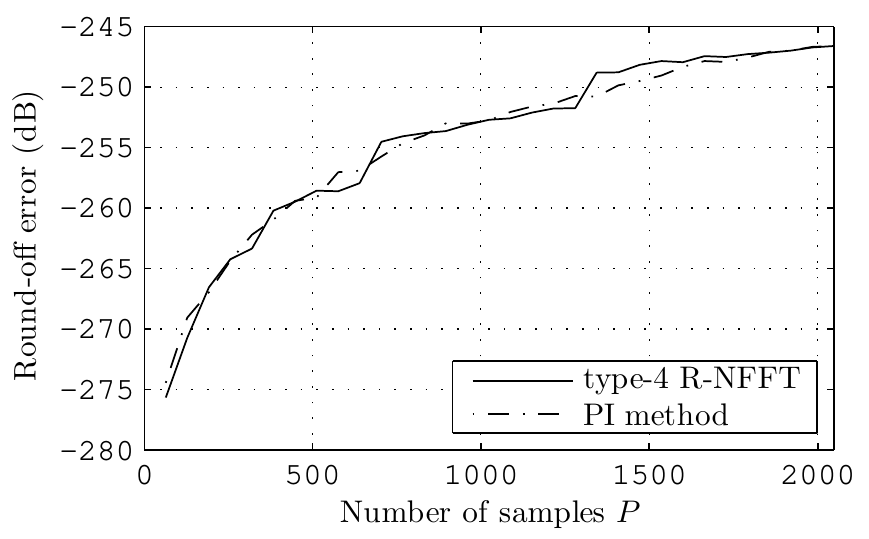}}
  \caption{\label{fig:3} Round-off error versus number of samples $P$ for the 
    type-4 R-NFFT method. }
\end{figure}

\Fig{
  \put(0,0){\includegraphics{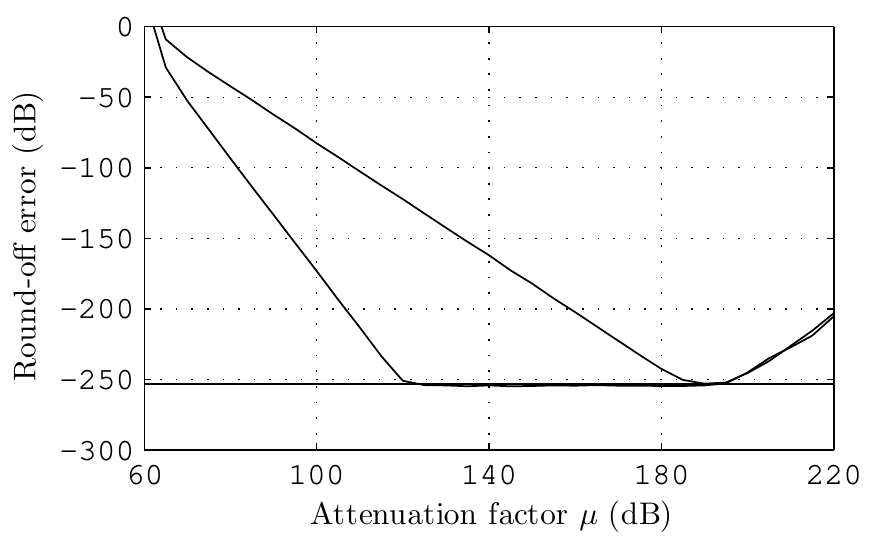}}
  \put(4.8,3.1){$\eta=1$}
  \put(3.7,2.2){$\eta=2$}
  \put(1.9,1.3){GE,CG methods}
}{
\label{fig:2} Round-off error versus attenuation factor ($\mu$) for two
oversampling factors ($\eta$) for the iterated type-4 NFFT method.}{0}
Fig \ref{fig:2} shows the round-off error of the type-4 R-NFFT method versus the
attenuation factor $\mu$ for $\eta=1$, 2. Note that without any oversampling ($\eta=1$),
we may select an attenuation $\mu$ for which its round-off error is similar to that of the
GE or CG methods. Fig. (\ref{fig:3}) shows the same round-off error but versus $P$ for
$\eta=1$. For each abscissa in this figure, $\mu$ has been selected to minimize the
round-off error. We can see that the type-4 R-NFFT method reaches the GE, CG benchmark for
typical $P$ values.

Finally, Fig. \ref{fig:6} shows the flop count of four methods: CG, GE, type-4 NFFT with
$\eta=6$, and type-4 R-NFFT with $\eta=1$. Note that the last two methods roughly have the
same flop count but, as shown in Fig. \ref{fig:1}, the NFFT method is slightly above the
GE, CG benchmark, while the R-NFFT method reaches it (Fig. \ref{fig:3}).
\Fig{
\put(0,0){\includegraphics{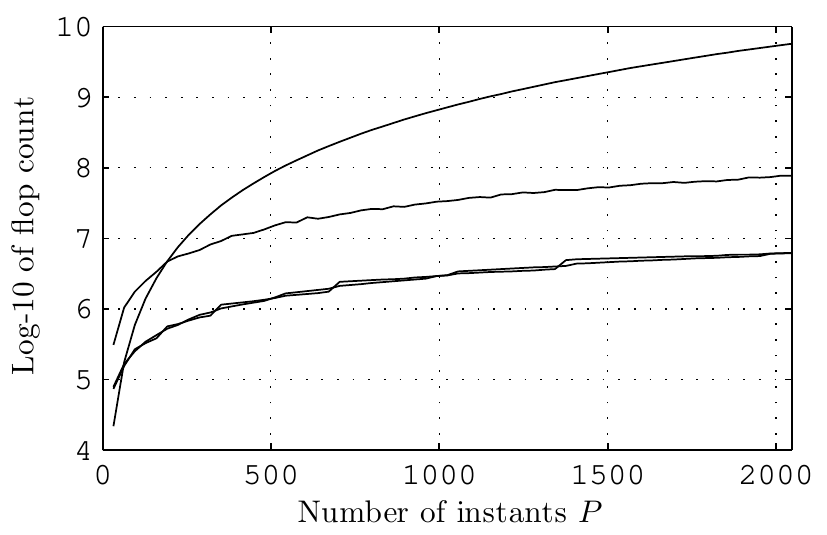}}
\put(3.5,2.2){NFFT $\eta=6$, R-NFFT $\eta=1$}
\put(5,3.65){CG method}
\put(1.7,4){GE method}
}{
\label{fig:6} Flop counts of four methods: GE, CG, type-4 NFFT with $\eta=6$, and
    type-4 R-NFFT with $\eta=1$. }{0}

\section{Conclusions}

We have presented non-iterative methods for the type 4 and 5 NFFTs in the one-dimensional
case, which are significantly less expensive computationally that the state-of-the-art
methods like the Conjugate Gradient. The methods are based on expressing the Lagrange
formula in terms of nonuniform convolutions that can be efficiently evaluated using the
type 1 and 2 NFFTs. The paper contains several numerical examples in which the proposed
methods are compared with the Gaussian elimination (GE) and conjugate gradient (CG)
methods in terms of round-off error and computational burden.

\bibliographystyle{IEEEbib}

\bibliography{../../../Utilities/LaTeX/Bibliography}

\end{document}